\documentstyle[preprint,eqsecnum,aps]{revtex}
\begin{document}
\tightenlines 
\draft

\title{A two-pion exchange three-nucleon force based on a realistic $\pi N$
interaction}
 
\author{T.-Y. Saito$^1$ and J. Haidenbauer$^2$}
 
\address{$^1$International Buddhist University, Habikino, Osaka 583, Japan}

\address{$^2$Institut f\"{u}r Kernphysik, Forschungszentrum J\"{u}lich
GmbH, D--52425 J\"{u}lich, Germany}
 
\maketitle
 
\begin{abstract}
The contribution of a $\pi \pi$-exchange three-body force to the 
three-nucleon binding energy is calculated in terms of a $\pi N$ amplitude. 
The latter is based on a meson-theoretical model of $\pi N$ interaction
developed by the J\"ulich group. Similar to a previous study based on
simple phenomenological $\pi N$ potentials a very small effect of the resulting
three-body force is found. Possible origins of the two-orders-of-magnitude
descrepancy between the present result and the values obtained for the
Tucson-Melbourne three-body force are investigated. Evidence is provided
that this discrepancy is most likely due to strikingly different off-shell
properties of the $\pi N$ amplitudes underlying the two approaches. 
\end{abstract}
\noindent

\section{Introduction}

Recently T.-Y. Saito and I. R. Afnan (SA) \cite{SA94,SA95} have calculated 
the contribution of a $\pi \pi$-exchange 
three-body force (see Fig. \ref{ftbf}) to the 
three-nucleon binding energy in terms of the $\pi N$ amplitude using 
perturbation theory. Their approach determines the contributions of the
different $\pi N$ partial waves and (via the division of the
$\pi N$ amplitude into a pole and nonpole term) allows for a consistent
determination of the $\pi NN$ form factor. The calculations are
based on phenomenological separable $\pi N$ potentials \cite{McLeod}.
The total contribution of this three-body force (TBF) to the binding
energy of the triton has been found to be very small. It is typically
of the order of a few keV. This result falls short of 
calculations based on the Tucson-Melbourne (TM) and the Brazilian
$\pi \pi$-exchange three-nucleon potentials \cite{TM1,TM2,Braz}.
The latter potentials make a contribution to the binding energy of the triton
that is of the order of the discrepancy between experiment and
calculations with realistic nucleon-nucleon potentials (i. e. about 1 MeV).

The origin of this surprisingly large difference of
two orders of magnitude in the contribution of the three-body force 
has been the topic of two subsequent 
very detailed investigations by Saito and Afnan \cite{SA95} and Murphy
and Coon \cite{MuCo}. SA concluded that the total contribution of the TBF
to the binding energy of the triton found in their approach is so 
small as a result of the energy dependence of the $\pi N$ amplitudes, 
cancellations between the contributions from the S- and P-wave $\pi N$
partial waves, and in particular, the soft $\pi NN$ form factor. 
Indeed the form factors extracted from their $\pi N$ interaction models
correspond to monopole cutoff masses of around or even less than 0.4 GeV - 
which have to 
be compared to values of about 0.7 - 0.8 GeV suggested from other 
information \cite{Liu,Coon90,ThoHo} and to the value of 0.8 GeV used in 
calculations with the TM potential.

Murphy and Coon carried out a thorough comparison of the $\pi N$ amplitudes
that underlie the TM TBF and the calculations of SA \cite{MuCo}.
They criticized that the amplitudes used by SA do not fulfil the
low-energy theorems of the $\pi N$ interaction as imposed by chiral
symmetry. But still they attested that these amplitudes are qualitatively
similar to the one used in the TM potential. The prime reason for the
two-orders-of-magnitude discrepancy found in the contributions of the
TBF was suspected to be likewise the soft $\pi NN$ form factor emerging
from the phenomenological separable $\pi N$ interactions employed by SA. 

In the present paper we want to re-investigate the origin of this large
discrepancy in the predicted contribution of the TBF. We 
follow the same approach as SA. However, we start out from a 
meson-theoretical $\pi N$ model developed recently by the J\"ulich 
group \cite{CSch1}. This model, besides being conceptionally much
better founded than the simple separable potentials employed in Refs. 
\cite{SA94,SA95}, has the important advantage that it does not exhibit
those deficiencies which led to a criticism on the work by Saito and 
Afnan. Firstly, the J\"ulich 
model is in agreement with empirical information on the $\pi N$
amplitude in the subthreshold region \cite{Hoff}. In particular its 
prediction for the amplitude at the so-called Cheng-Dashen point is 
close to the empirical value. Second, and most importantly it yields 
$\pi NN$ form factors which are comparable to those of the TM potential, 
with monopole cutoff masses of around 0.7 GeV \cite{CSch2}.
Therefore this model provides an ideal starting point for re-assessing
the role of the $\pi \pi$-exchange TBF in the binding of the
three-nucleon system. 

The paper is structured in the following way: In Section II we review
the salient features of the J\"ulich $\pi N$ model. In particular we
concentrate on those properties that are relevant for the present
study. The specific structure of our three-nucleon code makes it 
necessary to represent the meson-theoretical $\pi N$ amplitude in 
separable form. For this purpose
we applied the so-called EST method \cite{EST} which allows to generate
separable representations that agree exactly (on- and half-off-shell)
with the original interaction at specific predetermined energies.
This method is briefly described also in Section II. Furthermore, we discuss 
the reliability of the separable representation by comparing (off-shell)
amplitudes obtained from it to the ones of
the original J\"ulich model for various $\pi N$ partial waves. 

The formulation of the TBF is given in Section III together 
with a short outline of the formalism. Results for the contribution of
the $\pi \pi$-exchange TBF to the triton binding energy, based on the
$\pi N$ amplitude of the J\"ulich model are presented in Section IV. 
Anticipating our results we find again that the TBF is
very small. Therefore, in Section V, we embark on a detailed discussion of
the $\pi N$ amplitudes on which the TM TBF and our calculation are based. We
focus specifically on the off-shell properties of these amplitudes since
they are determined quite differently in the two approaches. Indeed, 
we will argue that much of the observed two-orders-of-magnitude discrepancy
in the binding energy is due to off-shell effects and we will present numerical 
evidence for this conjecture. Finally, a summary is given in Section VI.

\section{The $\pi N$ interaction}

The $\pi N$ models employed in the present investigation are based
on meson exchange and have been developed by the J\"ulich Group
\cite{CSch1}. They include the s-channel and u-channel nucleon and
delta-isobar pole diagrams together with correlated $\pi\pi$
exchange 
in the $J^P = 0^+$ ($\sigma$) and $1^-$ ($\rho$) channels
as shown in Fig. \ref{fig:diags}. 
The interaction potentials are derived in time ordered perturbation theory
and then unitarized by means of a relativistic (Lippmann-Schwinger type)
scattering equation
\begin{equation}
T \ = \ V \ + \ V G_0 T \ .
\label{ls}
\end{equation}
The resulting models account for the scattering data
in the elastic region as well as for the low-energy parameters
\cite{CSch1}.
Furthermore they also satisfy chiral symmetry constraints.
In particular, the resulting values for the so-called $\pi N$ $\Sigma$
term (the contribution of the isoscalar forward scattering
amplitude at the Cheng-Dashen point) - $\Sigma$ = 66.4 (65.6) MeV
for model 1 (2) of Ref. \cite{CSch1} -
are in good agreement with the empirical value of $\Sigma$ = 60 MeV
\cite{GL}.
 
Since in the discussion of Murphy and Coon \cite{MuCo} special
attention was given to the subthreshold behaviour of the $\pi N$
amplitudes, we want to present here the corresponding results for the J\"ulich
$\pi N$ model. In the continuation of the T-matrix for the potential models
to the subthreshold region, we follow the procedure outlined in section 4 of 
Ref. \cite{MuCo}. We calculated the (on-shell) background isoscalar amplitude
$\bar F^+(\nu, t)$ (conventionally called $\bar D^+$; for definition see, 
e.g., Refs. \cite{MuCo,Hoe}) from the subthreshold point $\nu = 0$, $t=0$ to
the Cheng-Dashen point ($\nu = 0$, $t=2m^2_\pi$). The predictions of the 
J\"ulich models 1 and 2 are shown in Fig. \ref{fig:cdp} in comparison to the 
ones of the $\pi N$ amplitudes that form the basis of the TM and Brazil 
$\pi \pi$-exchange TBF. One can see that the results of the meson-exchange 
models more or less coincide with the $\pi N$ amplitude employed in the TM 
TBF. Furthermore, they are also in rather nice agreement
with the empirical subthreshold amplitude given by H\"ohler \cite{Hoe}. 

In a recent paper C. Sch\"utz et al. \cite{CSch2} have determined the
$\pi NN$ vertex functions resulting from the models of Ref. \cite{CSch1} 
(cf. Ref. \cite{CSch2} and the Appendix for definitions and relevant
formulue). 
From the vertex functions at the nucleon pole, the decrease in the $\pi NN$
form factors $F_{\pi NN} (p^2)$ from the pion pole $p^2=m_\pi^2$
to $p^2=0$, has been extracted following the procedure proposed by 
Mizutani et al. \cite{Mizu}. This quantity is a measure for
the softness of the $\pi NN$ form factor.
It has been found that the $\pi NN$ form factors implied by the models 
considered in Ref. \cite{CSch2} are, in general, significantly 
harder than the ones used by SA in their study of the contribution
of the $\pi\pi$-exchange three-nucleon force to the $3N$ binding
energy. In particular, the model 2' of Ref. \cite{CSch2} yields a
decrease in the form factor from the pion pole to $p^2=0$ 
of slightly less then 4\% - which is quite close to the
value of 3\% implied by the form factor introduced in the TM
three-nucleon force and also in good agreement with a recent
lattice QCD calculation \cite{Liu} and other independent
information \cite{Coon90,ThoHo}. Accordingly, the major concern
raised by Murphy and Coon against the work of SA does not apply
for this model and therefore it provides an excellent starting
point for re-analyzing the contributions of the
$\pi\pi$-exchange three-nucleon force to the $3N$ binding
energy in the approach of Saito and Afnan.

In this work we will also present results for the other models
considered in Ref. \cite{CSch2} and it is appropriate to say
some words about the basic differences between these models.
All the models are based on the same dynamical input (cf. Fig.
\ref{fig:diags}). They differ, however, in the (phenomenological)
parametrization of the (bare) vertex form factors. 
Details and explicit formulae can be
found in sect. III of Ref. \cite{CSch1}. Both models provide a
similarly good description of the $\pi N$ scattering data.
However, the differences in the parametrization of the vertex
form factors lead to different vertex functions and in turn to
different (dressed) $\pi NN$ form factors.

Two further models (1' and 2') have been presented in Ref.
\cite{CSch2} for the following reason: The models 1 and 2 are
constructed by assuming $\pi NN$ pseudovector ($pv$) coupling.
However, the $\pi NN$ form factor derived from 
lattice QCD calculations \cite{Liu}, and the values of $F_{\pi NN} (0)$
from other independent
information \cite{Coon90,ThoHo} (and also the values given
for the models used by SA), are based on pseudoscalar ($ps$)
coupling. In order to allow for a meaningful comparison for the
different couplings the authors of Ref. \cite{CSch2} 
have constructed variants of models 1 and 2 (labelled
1' and 2') where $ps$ coupling is used in the nucleon s-channel pole terms.

Values for $F_{\pi NN}(0)$ for the various models are compiled
in Table I. (Note that $F_{\pi NN}$ is normalized to
$F_{\pi NN}(m_\pi^2)=1$.) For the ease of comparison we also
include here the result for the separable $\pi N$ model $PJ$ by
McLeod and Afnan \cite{McLeod} which has been used (amongst others)
in the investigation by SA and on which most of the concerns and criticism
of Murphy and Coon are based. It is evident that for this model the 
decrease of the form factor from the pion pole to $p^2=0$ is 
much larger - almost 20\%.
We want to emphasize here, however, that it is the $\pi NN$ vertex
function which enters into the calculation of the TBF and not the form 
factor (cf. Section III). Therefore we also show this
quantity (cf. Fig. \ref{fig:fofa}). Note that the $\pi NN$ vertex functions
for models 1 and 1' are practically the same. Accordingly we expect these
models to give the same result for the contributions of the
$\pi\pi$-exchange TBF to the $3N$ binding energy, 
and therefore we will consider only model 1 in the
following analysis. Furthermore, the vertex function for model 2 is
obviously harder than the one resulting from model 2' (for small
and intermediate momenta) - in contrast the form factors
for model 2' looks harder (cf. Table I). This seemingly
paradoxical situation has been discussed thoroughly in Ref.
\cite{CSch2}.

Before carrying out the actual three-nucleon calculations
the meson-theoretical $\pi N$ models have to be expanded in
separable form. This is necessitated by the specific structure
of our three-nucleon code which can only deal with
interaction models given in separable form. For this purpose
we apply the so-called Ernst-Shakin-Thaler (EST) method \cite{EST}
which allows us to generate separable representations of arbitrary
rank $N$ that agree exactly (on- and half-off-shell)
with the original reaction matrix at $N$ specific predetermined
energies.

Let us begin with the (partial wave projected) Lippmann-Schwinger equation 
for the radial wave function  
\begin{equation}
|\psi_E \rangle = |k_E\rangle + G_0 (E)
V |\psi_E\rangle,
\label{est1}
\end{equation}
where $|k_E\rangle$ is the incoming wave and $G_0(E)$ the two-body  
Green's function. (In Eq. (\ref{est1}) and in the following the partial 
wave index is suppressed for convenience.) 
For proper scattering solutions (on-shell) $k_E$ and $E$ are related by 
\begin{equation}
E = \sqrt{m_N^2 + k^2_E} + \sqrt{m_\pi^2 + k^2_E}. 
\label{engy}
\end{equation}
According to the EST method a rank-$N$ separable representation for the 
potential $V$ is given by
\begin{equation}
\tilde V = \sum_{i,j=1}^N V|\psi_{E_i} \rangle \lambda_{ij} 
\langle \psi_{E_j} | V \ ,
\label{est2}
\end{equation}
where $E_i$, ($i=1,...,N$), are $N$ freely chosen energies.
The coupling strengths $\lambda_{ij}$ are determined by the condition
\begin{equation}
\sum_{j=1}^N \lambda_{ij} \langle \psi_{E_j} | V  
|\psi_{E_k} \rangle = \delta_{ik} \ . 
\label{est3}
\end{equation}
It is evident from Eq. (\ref{est2}) that the "form factors" of the 
separable potential $\tilde V$ \cite{Com} consist of the objects 
$V|\psi_{E_i} \rangle$, where $|\psi_{E_i}\rangle $ are solutions of 
Eq. ({\ref{est1}) for the potential $V$ at the energies $E_i$. Therefore, 
by virtue of Eq. (\ref{est3}), the following relation holds
\begin{equation}
\tilde V |\psi_{E_i} \rangle = V |\psi_{E_i} \rangle = 
 T (E_i) |k_{E_i} \rangle =  \tilde T (E_i) |k_{E_i} \rangle  
\label{est4}
\end{equation}
at the $N$ energies $E_i$, where $\tilde T$ is the solution of the
Lippmann-Schwinger equation (\ref{ls}) for the separable representation 
$\tilde V$. This means that the on-shell as well as the 
half-off-shell t-matrix for both interactions $V$ and $\tilde V$ are exactly
the same at the energies $E_i$. 

In the present case the interaction models $V$ are energy-dependent and 
therefore a modification of this scheme proposed by B. Pearce \cite{Pearce}
is employed. According to it the condition
\begin{equation}
\langle \psi_{E_l} | V (E) |\psi_{E_k} \rangle  = 
\langle \psi_{E_l} | \tilde V (E) |\psi_{E_k} \rangle = 
\sum_{i,j=1}^N 
\langle \psi_{E_l} | V (E_i) |\psi_{E_i} \rangle  
\lambda_{ij} (E)
\langle \psi_{E_j} | V (E_j) |\psi_{E_k} \rangle  
\label{est5}
\end{equation}
has to be used for determining the coupling strengths $\lambda_{ij}(E)$ 
instead of Eq. (\ref{est3}). As expected also the separable representation 
becomes now energy-dependent. 

A special treatment is required for the $P_{11}$ partial wave
which contains the (s-channel) nucleon pole. It must be possible
to clearly separate the contribution of this pole term from the
total $P_{11}$ amplitude. Its contribution to the three-nucleon
binding energy is already taken into account by solving the
standard bound-state Faddeev equations. Therefore, in order to
avoid double counting, only the non-pole part of the $P_{11}$
must be considered in the present investigation. Furthermore in
the consistent approach of SA the $\pi NN$ vertex function is
extracted from this pole term and is then used for the vertices
where the pions are emitted (absorbed) by (at) the outer nucleons.
Consequently a separable representation for the $P_{11}$ partial
wave must guarantee that (a) the non-pole amplitude is reliably
reproduced and (b) the $\pi NN$ vertex function extracted from
the pole term agrees exactly with the one obtained for the
original interaction model. This can be achieved and we summarize details 
of the construction procedure in the Appendix.

Of course if one relies on such separable representations one has
to ensure that they incorporate all the relevant properties
of the original interaction models. From extensive tests, we find 
that a rank-1 separable representation is sufficient for
the present purpose provided that the expansion energy is
choosen in the relevant energy domain (i. e. around the
triton binding energy which corresponds to a center-of-mass
energy of roughly 930 MeV in the $\pi N$ system). Thus we have
selected this particular energy for the separable representation
to be applied in the present study. Since this energy is below the elastic
$\pi N$ threshold, $k_E$ in Eq. (\ref{est1}) can no longer be fixed by
the on-shell condition. Following our previous work \cite{HK86} we choose
$k_E$ in such a way that $ik_E$ fulfils Eq. (\ref{engy}). For 
$E_1 = 930$ MeV this implies $k_{E_1} \approx 138$ MeV/c. 

The quality of the separable represention is demonstrated in the
Figs. \ref{fig:pin1} and \ref{fig:pin2} for model 2. Fig. \ref{fig:pin1}
shows the off-shell transition amplitude
$T_\alpha (q,q';Z)$ for fixed off-shell momenta $q  = q'  =  130$ MeV
as a function of the total energy $Z$. Note that in case of
the $P_{11}$ partial wave only the non-pole part of the t-matrix is shown
since, as explained above, this is the part relevant for the
present study. Fig. \ref{fig:pin2} shows the transition amplitude
$T_\alpha (q,q';Z)$ for fixed $Z$ and $q'$ as a function of the other
off-shell momentum. In the latter figure we display only results
for those partial waves that are expected to give the dominant
contribution of the $\pi\pi$-exchange three-nucleon force to the $3N$
binding energy \cite{SA94}, namely $S_{31}$, $P_{11}$, and $P_{33}$.
We want to point out, however, that the agreement in the
other partial waves is of similar quality. Likewise we want to 
refrain from displaying corresponding results for the model 1 here
since the quality of its separable representation is pretty much the same.
Furthermore, we do not show the $\pi NN$ vertex functions resulting from the
separable representation because - by construction - they are
identical to the ones of the original model (cf. the Appendix).

Note that we have also constructed rank-1 separable representations
where the expansion energies are near the $\pi N$ threshold
($E_1  =  1077$ MeV for $S_{11}$; $E_1  =  1000$ MeV for $S_{31}$; 
$E_1  =  1100$ MeV for $P_{11}$, $P_{31}$, $P_{13}$,
$P_{33}$). 
(In case of the $S_{31}$ partial wave the amplitude around 1077 MeV
has a peculiar energy dependence which would require, in principle, a
rank-2 representation. In order to avoid this complication we have 
chosen a  somewhat lower value for $E_1$.)
These representations will be also employed 
in our investigations. They will serve as a term of reference for how
strongly the resulting $3N$ binding energies depend on the
specific choice of the separable representation.

\section{Formulation of the three-body force}

In this section we formulate the TBF using the
$\pi N$ amplitude and $\pi NN$ vertex function which have been
derived in the previous section.

We follow the prescription by SA. The TBF is
schematically shown in Fig.~\ref{figtbf1}, i.e., (i) a pion
is emitted from the first nucleon, (ii) the pion is scattered
off the second nucleon, (iii) the pion is absorbed
by the third nucleon. The strength function of pion emission
and absorption is given by the $\pi NN$ vertex function which
depends on the energy of the $\pi N$ system as explained
in the previous section. Also the $\pi N$ amplitude corresponding to the
scattering of the pion on the second nucleon is energy dependent and
includes only the non-pole contribution in the $P_{11}$ channel. We
introduce Jacobi variables in the $\pi NNN$ system so that
we define the $\pi N$ relative momenta and energies at the
stages (i), (ii) and (iii).

The momenta of the three nucleons before the three-body interaction
in Fig.~\ref{figtbf1} are given by ${\bf k}_1$, ${\bf k}_2$
and ${\bf k}_3$ and the momenta after the interaction are
${\bf k}'_1$, ${\bf k}'_2$ and ${\bf k}'_3$, respectively.
We define the relative momenta between the third nucleon
and the system of first and second nucleons and pion (${\bf q}_3$),
between the second nucleon and the system of first nucleon and pion
(${\bf p}_3$) and between the first nucleon and pion (${\bf Q}_3$)
at the stage (i)
\begin{eqnarray}
{\bf q}_3 &=& -{\bf k}_3,
\label{tbf11}  \\
{\bf p}_3 &=& {m_N ({\bf k}_\pi + {\bf k}'_1)
- (m_N + m_\pi) {\bf k}_2 \over
(2m_N + m_\pi) },
\label{tbf12}  \\
{\bf Q}_3 &=& {m_N {\bf k}_\pi - m_\pi {\bf k}'_1 \over
(m_N + m_\pi)},
\label{tbf13}
\end{eqnarray}
where ${\bf k}_\pi$ is the pion momentum. Then the center-of-mass energy in the
system of first nucleon and pion, $E_3$, is obtained
using those relative momenta
\begin{equation}
E_3 = E + m_N - {q_3^2 \over 2\mu_2} - {p_3^2 \over 2\mu_1}
\label{tbf14}
\end{equation}
where $E = -E_T$ is the total energy of the whole system not
including rest masses, and the reduced masses $\mu_1$ and
$\mu_2$ are defined, respectively, by the relations
\begin{equation}
{1\over \mu_1} = {1\over m_N} + {1\over m_N + m_\pi},
\label{tbf15}
\end{equation}
and
\begin{equation}
{1\over \mu_2} = {1 \over m_N} + {1 \over 2m_N + m_\pi}.
\end{equation}
In the same way we define the relative momenta and energies
of the system of second nucleon and pion,
${\bf Q}'_3$ (before scattering), ${\bf Q}'_1$
(after scattering) and $E_2$ at the stage (ii) and ${\bf Q}_1$
and $E_1$ of the system of third nucleon
and pion at the stage (iii), respectively.

Using these $\pi N$ relative momenta and energies
the TBF, $W(E)$, can be symbolically written as 
\begin{equation}
W(E) = v^R_{\pi N}(Q_1; E_1) \ G_{\pi NNN}(E)
\ T_{\pi N}(Q'_1, Q'_3; E_2) \ G_{\pi NNN}(E)
\ v^R_{\pi N}(Q_3; E_3).
\label{tbf1}
\end{equation}
Here, $v^R_{\pi N}(Q_i; E_i)$ ($i =$ 1, 3) is the renormalized $\pi NN$ vertex 
function (cf. the Appendix) which gives the strength of the pion emission and
absorption on the nucleon. $G_{\pi NNN}(E)$ is the propagator
of the $\pi NNN$ system and $T_{\pi N}(Q'_1, Q'_3; E_2)$
is the non-pole part of the $\pi N$ scattering amplitude. Eq. (\ref{tbf1})
represents how the pion is emitted from the first nucleon, 
scattered off the second nucleon, and then absorbed by the
third nucleon. Note that the $\pi N$ amplitude and the
$\pi NN$ vertex function are determined in the same framework, i. e. 
they are obtained from the same $\pi N$ interaction model. A detailed 
discussion of Eq. (\ref{tbf1}) can be found in Ref. \cite{SA95}.
Since in the J\"ulich model the energy is defined fully relativistic
whereas a semi-relativistic form is employed by SA, we have 
to change the $\pi NNN$ propagator of Ref. \cite{SA95} to be
\begin{equation}
G_{\pi NNN} (E) = \Bigl(E_i - \sqrt{Q_i^2 + m_N^2}
 - \sqrt{Q_i^2 + m_\pi^2} \Bigr)^{-1},
\label{thb2}
\end{equation}
with $i =$ 1 or 3  (cf. Eq. (4.9) of Ref. \cite{SA95}).


\section{Results}

The contribution of the TBF to the binding energy of the
three-nucleon system, $\Delta E^{(3)}$, is calculated in first order
perturbation theory, i. e. 
\begin{equation}
\Delta E^{(3)} = \langle\Psi|W(-E_T)|\Psi\rangle,
\label{rst1}
\end{equation}
where $|\Psi\rangle$ is the triton wave function. This wave function is
obtained from solving the Faddeev equations for the 
so-called PEST potential \cite{HK86,KHP87} which is a
separable representation of the Paris $NN$ potential \cite{paris}
derived by the EST method.
All nucleon-nucleon partial waves with total angular momentum less than or
equal two are employed in the calculation. The 
triton properties obtained by the PEST potential are comparable
with those by the original Paris potential as shown in Refs.
\cite{Parke,SA95}.

Evidently, like in the work by SA there is no consistency between the
$NN$ interaction which is used to generate the triton wave function 
and the TBF. However, we are making use of the Born 
approximation and therefore the triton wave function and the TBF
are obtained separately, anyway. Thus, as argued already by SA \cite{SA95},
we do not expect that this inconsistency has an influence on the qualitative
features of our results. One should also keep in mind that a similar 
inconsistency is involved in standard $3N$ calculations employing
the TM TBF. 

The contributions of the TBF generated by the $\pi N$ models 1, 
2 and 2' of Ref. \cite{CSch2} to the $3N$ binding energy 
are listed in Table \ref{rstt11}. 
These results are based on the separable representations  
of the original interaction models described in the preceeding section. 
Note that the value of 930 MeV is used for the expansion energy 
since we expect the average $\pi N$ energy in the three-nucleon system to 
be near this value. 
In Table \ref{rstt11} also the result for model $PJ$ of Ref.
\cite{SA95} is listed for the sake of comparison. Thus, it is easy to see 
that our results are qualitatively very similar to those of SA. Again the
contibutions of the individual $\pi N$ partial waves to the $3N$ binding 
energy are very small (only in the order of
$keV$) and there is again a strong cancellation between the
attractive $P_{11}$ and $P_{33}$ partial waves and the
repulsive $S_{31}$ partial wave. 

In Ref. \cite{MuCo} Murphy and Coon have presented a thorough comparison
of the $\pi N$ amplitudes used in the TM and Brazilian TBF
and the one applied in the calculations by SA. Their main conclusion was
that the two-orders-of-magnitude smaller value for the binding energy
resulting from the $\pi \pi$-exchange
TBF based upon the separable models of SA is most likely due to
the very soft form factor resulting from these potentials. Therefore we
want to look now at the $\pi N$ form factors extracted from
the models applied in the present study. The values for $F_{\pi NN}(0)$
are also given in Table \ref{rstt11}. Note
that model 2' yields the hardest $\pi NN$ form factor which corresponds to
a monopole form factor with a cutoff mass of about 708 MeV whereas
the model $PJ$ of SA corresponds to a monopole form factor with a cutoff 
mass of just 317 MeV. From comparing different columns of Table \ref{rstt11}
one might conclude that there is some influence of the form factor on the 
magnitude of the TBF. However, the
variations in the individual partial wave contributions are only around a 
factor of two or three and definitely not two orders of magnitude. 
Furthermore, the cancellation effects are independent of the softness of
the $\pi NN$ form factor and they tend to reduce the variations in the 
total contribution. In fact, the results for the $\pi N$ models considered
in the present paper lie all within a range of 15 $keV$, as can be seen
from Table \ref{rstt11}. 

At this point one may wonder how reliable results based on a rank 1
separable approximation of the J\"ulich models are. 
In order to estimate the uncertainty due to the simplicity of the
representation we constructed another rank 1 represention where the
expansion energy was chosen at $\pi N$ threshold (for the S-waves)
or slightly above. Specifically we chose 1077
MeV for $S_{11}$, 1000 MeV for $S_{31}$ and 1100 MeV for the other
partial waves.  
The results for these alternative separable representations are compared
with the ones obtained for our 'standard' choice in Table \ref{rstt2}.
From this Table we see that there are variations of the order of $20\sim30\%$
in some partial waves - but qualitatively there is no change in the
results. Therefore we are confident that the used rank-1 separable
representations are sufficiently accurate for the aim of the present
investigation. 


\section{Discussion}

We conclude from the previous section that the softness of
the $\pi NN$ form factor is not responsible for the smallness
of the $\pi\pi$-exchange TBF based on $\pi N$ potential models. 
If so, what makes the contribution to the three-body binding energy so small? 
In order to shed some light on this let us examine
the basic two differences between the TM TBF and
the one derived from a $\pi N$ potential model. The first difference
concerns the energy dependence. The original $\pi N$ amplitude on
which the TM TBF is based, is given in a covariant form and therefore
depends on the energy (of the pion). However, in order to make this TBF
suitable for application in standard (non-relativistic) 3N calculations
the $\pi N$ amplitude is expanded in powers of $k_\pi/m_N$ and only the
lowest order terms are kept. As a consequene the initial energy dependence 
drops completely out of the resulting TBF. In the approach
of SA the TBF is energy dependent in a two-fold way, 
namely via the $\pi N$ amplitude but also via the $\pi NN$ form factor. 
The effect of switching off this energy dependence in the TBF
has been analyzed thoroughly in Ref. \cite{SA94,SA95} where it was found
that it leads to a sizeable increase in the resulting binding energy. 
However, it was concluded by SA that the approximation of fixing the
energy in the $\pi N$ amplitude and the $\pi NN$ form factor does not
lead to a sufficiently large change in the contributions to make them
comparable with the result for the TM TBF. 
None the less we would like to look at this point again because  
(unlike the potentials employed by SA) now the $\pi N$ interaction model 
itself is energy dependent and therefore the effects from the energy
dependence might be stronger. Note that in the present case the
energy dependence of the TBF enters at three levels:
(a) the energy dependence of the $\pi N$ potential itself;
(b) the energy dependence of the $\pi N$ t-matrix;
(c) the energy dependence of the $\pi NN$ vertex function.

We investigate the role of the energy dependence by fixing the energy at
$E = 930$ MeV. This is the value chosen for constructing the 
separable interaction via the EST method and, accordingly, where the  
$\pi N$ amplitudes generated by the J\"ulich model and its separable
representation agree almost exactly. Thus, the results for our rank-1
separable interaction should be practically identical to the one for
the original J\"ulich model in the particular case where the whole energy 
dependence (a)-(c) is fixed. Corresponding values are given in the last 
column of Table \ref{dsct3}. The numbers in the third column of Table
\ref{dsct3} are obtained by fixing only the energy dependence of the
$\pi N$ interaction. For the ease of comparison results 
without restricting the energy dependence are also shown in the Table. 
We see that the contribution of each $\pi N$ partial wave
is enhanced by about 50\% after fixing the energy dependence of the
potential. Fixing the energy dependence also in the $\pi N$ t-matrix 
and the $\pi NN$ form factor leads to a further increase in the 
contributions from the $P_{11}$ and $P_{33}$ waves 
and to a slight suppression in the $S_{31}$. However, those approximations 
definitely do not provide any really substantial enhancement of the resulting 
triton binding - in line with the findings of SA. 
All results shown in Table \ref{dsct3} are obtained by using model 2'.
The other models behave qualitatively very similar and therefore we don't
give the corresponding numbers here. 

The other major difference between the TM force and the TBF based on a
$\pi N$ potential concerns the off-shell extrapolation of the $\pi N$
amplitude. In the J\"ulich model the off-shell properties of the $\pi N$
amplitude are completely determined by the dynamical ingredients of the
$\pi N$ model and the fit to the $\pi N$ data. Since also for the potential 
models employed by SA the off-shell properties are, in principle, constrained 
by a fit to $\pi N$ data let us emphasize the main difference here. 
In case of phenomenological separable interactions the off-shell behavior is, 
to a large extend,  determined by the specific choice of the form-factor
function. Moreover, separable interactions act only in single partial waves 
and accordingly the free parameters are determined by fitting only a
single partial wave. On the other hand, in a meson-exchange potential like
the J\"ulich model the dynamical ingredients give, in general, contributions 
to all partial waves and therefore the free parameters in this model - which 
are essentially the cutoff masses in the (baryon-baryon-meson) vertex form
factors \cite{CSch1} - are much better constrained by a fit to the $\pi N$ data.
Clearly also here differences in the dynamics and/or differences in the
parametrization of the vertex form factors will lead to variations in the
off-shell properties of the resulting $\pi N$ amplitude. Indeed such
differences exist between the models 1 and 2 (or 2') considered here. But, 
as we have already seen in the last section, they to not lead to any 
significant variations in the results for the TBF. 

The $\pi N$ amplitude used in the TM force is given by 
(cf., e.g., Ref. \cite{MuCo,CoPe}) 
\begin{equation}
T^{ij}_{\pi N}( {\bf k}_\pi, {\bf k}_\pi ')
= F_{\pi NN}({\bf k}^2_\pi) F_{\pi NN}({\bf k}_\pi '^2)
\left\lbrace \delta^{ij} 
\left\lbrack a + b \ {\bf k}_\pi \cdot {\bf k}_\pi '
+ c \ ({\bf k}_\pi^2 + {\bf k}_\pi '^2) \right\rbrack  
-  d \ \epsilon^{ijk}\tau^k {\bf \sigma} \cdot 
{\bf k}_\pi \times {\bf k}_\pi ' \right\rbrace \ , 
\label{disc6}
\end{equation}
where $i,j$ are pion (cartesian) indices, 
${\bf k}_\pi$ and ${\bf k}_\pi '$ are the momenta of the incoming and
outgoing (off-shell) pions, and $a$, $b$, $c$, and $d$ are constants 
defined, e.g., in Ref. \cite{CoPe}.
Evidently the off-shell extrapolation is provided by the  
form factor function $F_{\pi NN}$ \cite{CoGl}. In the standard version of 
the TM force this form factors are assumed to be of monopole type, 
\begin{equation}
F_{\pi NN}({\bf k}_\pi^2) = {\Lambda_\pi^2 - m_\pi^2 \over
\Lambda_\pi^2 + {\bf k}_\pi^2},
\label{disc4}
\end{equation}
with a cutoff mass $\Lambda_\pi = 5.8 m_\pi \approx 800$ MeV. 
We would like to emphasize at this point that, in principle, there is no 
connection between the $\pi NN$ vertex (with an off-shell pion) and the $\pi N$ 
amplitude entering into the TBF. It would appear only in
the contribution of the direct nucleon pole diagram to the $\pi N$ amplitude,
which, however, is omitted in order to avoid double counting (cf. the 
discussion in section II). 
Therefore the prescription for the off-shell extrapolation used in the
TM force must be considered as rather arbitrary. 

In the discussion above we have tacitly ignored a conceptional subtlety
when we talk about "off-shell". In case of the J\"ulich model the
$\pi N$ amplitude is obtained off-energy-shell, while 
an off-mass-shell $\pi N$ amplitude is used in the TM potential. 
This means, that we can not make a simple comparison between them.
The off-pion-mass-shell $\pi N$ amplitude of the TM force depends on the 
pion momenta ${\bf k_\pi}$ and ${\bf k_\pi '}$.
The off-energy-shell value of the $\pi N$ amplitude for the potential
model (at a certain energy) is given as a function of the relative momentum
between the pion and the nucleon, ${\bf Q}$. Within non-relativistic 
kinematics ${\bf Q}$ is obtained by (cf. Eq. (\ref{tbf13})) 
\begin{equation}
{\bf Q} = { {\bf k}_\pi - \frac{m_\pi}{m_N} {\bf k}_N
\over 1 + \frac{m_\pi}{m_N} },
\label{disc3}
\end{equation}
where ${\bf k_N}$ is the nucleon momentum. One can see from this 
relation that ${\bf Q}$ becomes equivalent to
${\bf k_\pi}$ only in the limit of $m_\pi/m_N\rightarrow0$.

In the following we want to discuss the
off-shell properties entering into the calculations with the TM force
and into the results presented in this paper. In view of the aforementioned
difficulties it should be clear that any comparision can be only of 
qualitative nature. None the less, as we will see below such an analysis 
is useful and we believe that it indicates the source of 
the large discrepancy found in the contributions to the binding energy
from the two approaches. 
 
For this purpose let us introduce a half-off-shell function in the
following way,
\begin{equation}
f_\alpha (Q) = {T_\alpha (Q,Q';Z) \over  Q^l}  
\label{disc1}
\end{equation}
where $T_\alpha (Q,Q';Z)$ is the off-shell $\pi N$ t-matrix (projected on 
the partial wave $\alpha$ with angular momentum $l$) at a fixed energy $Z$ 
and a fixed momentum $Q'$. 
The factor $Q^l$ is taken out for convenience because then we can 
normalize these half-off-shell functions to $1$ at $Q=0$ for s- as well
p-waves and we can easily compare them with each other. 
Corresponding results for the J\"ulich $\pi N$ potential 2' are shown 
in Fig. \ref{figtbf2}, where $Z$ and $Q'$ in Eq. (\ref{disc1}) have 
been fixed to 930 MeV and 130 MeV/c, respectively. 

Let us first take a look at the p-waves and in particular
at the $P_{11}$ and $P_{33}$ partial waves which provide the
main attractive contributions to the TBF (cf. Table \ref{rstt11}).
In this case the momentum dependence of the TM $\pi N$ amplitudes 
is roughly given by a monopole type function $F({\bf Q}^2) = 
\Lambda^2 / (\Lambda^2+ {\bf Q}^2)$ with $\Lambda = 800$ MeV, 
cf. Eqs. (\ref{disc6}-\ref{disc4}), which is shown by the dashed curve
in Fig. \ref{figtbf2}. We observe that the corresponding 
half-off-shell functions of the J\"ulich $\pi N$ model 
fall off much faster with increasing (off-shell) momentum than
this function. Accordingly we expect that a TBF based on the 
potential model will yield a much smaller attractive 
contribution to the three-body binding than one with off-shell 
properties similar to the monopole type function. 
 
In case of the s-waves we get large repulsive contributions from
the $S_{31}$ partial waves and small attractive contributions from
$S_{11}$ (cf. Table \ref{rstt11}). Please note that also the 
corresponding half-off-shell functions are radically different. 
The one for $S_{31}$ exhibits a strong enhancement whereas the 
one for $S_{11}$ falls off very strongly with increasing (off-shell) 
momentum. The s-wave part of the TM force ($a$- and $c$-terms), on 
the other hand, has no isospin dependence, cf. Eq. (\ref{disc6}). 
This means that here the $S_{31}$ and $S_{11}$ partial waves
have exactly the same momentum dependence. Therefore we suspect that a 
strong cancellation between the contributions from those two s-waves 
takes place. Indeed, actual triton calculations
employing the TM force confirm that the total s-wave contributions 
are comparably small \cite{Kam}. Evidently, such a 
cancellation does not occur with the TBF based on the J\"ulich $\pi N$ 
model because of the differences in the off-shell properties.
As a consequence, the
(large) repulsive contribution of the $S_{31}$ partial wave
to the three-body binding survives (cf. Table \ref{rstt11}).

Summarizing this phenomenological discussion of the off-shell 
properties we expect that a calculation based on the
off-shell extrapolation used in the TM force should lead to a strong 
enhancement of the attractive contributions and at the
same time reduce the repulsive contributions.
We believe that this is the basic mechanism which makes the 
binding energy obtained with the TM TBF so large. 
We would like to 
substantiate this claim quantitatively with a model calculation. We can
do this by substituting the off-shell properties of our $\pi N$ model 
by the ones used in the TM force. This can be easily	
done for the separable representations that we are using. We only need
to replace the J\"ulich off-shell behavior by defining the form factor
of the separable potential, 
$g_\alpha (Q): = \langle Q | V_\alpha | \psi_{E_1} \rangle$, as follows: 
\begin{equation}
g_\alpha(Q) \rightarrow 
\Biggl(\lim_{Q\rightarrow0}{g_\alpha(Q)\over Q^l}\Biggr)
{Q^l\over1+Q^2/\Lambda^2} \ . 
\label{disc2}
\end{equation}
Furthermore we fix the energy dependence in the $\pi N$
amplitude and the $\pi NN$ vertex function again.

We demonstrate the effect on the binding energy for several values of 
$\Lambda$ in Table \ref{dsct4}.
From this Table it is clear that we can get a substantial increase in 
the triton binding energy by choosing $\Lambda\approx600\sim800$ MeV. 
Specifically one can see that the repulsion of the $S_{31}$ partial wave is 
suppressed and, moreover, cancels to a large extent with the $S_{11}$.
At the same time the attraction provided by
the $P_{11}$ and $P_{33}$ partial waves is strongly enhanced - as we
expected from analyzing Fig. \ref{figtbf2}. In fact, if we take into 
account that our results are based on first-order perturbation theory
and therefore may underestimate the correct values by a factor two or
even three \cite{CoPe} then our simulation with the choice $\Lambda = 800$ MeV 
practically reproduces the TM result, which is likewise based on
$\Lambda_\pi = 800$ MeV.

\section{Summary}

In the present paper we have re-investigated the origin of the large
discrepancy in the contribution of a $\pi\pi$-exchange TBF 
found by Saito and Afnan to the commenly accepted values obtained
with the Tucson-Melbourne or Brazil TBF. 
Unlike SA, who employed phenomenological separable potentials, we
started out from a meson-theoretical $\pi N$ interaction model developed
recently by the J\"ulich group. 
This model provides a good description of elastic $\pi N$ scattering
data. It is also in agreement with empirical information on the 
$\pi N$ amplitude in the subthreshold region. In particular,
it predicts the $\pi N$ $\Sigma$ term close to the empirical value.
Furthermore, the decrease in the $\pi NN$ form factors
$F_{\pi NN} (q^2)$ from $q^2=m_\pi^2$ to $q^2=0$ of about 4-7\% and 
is comparable to that of the TM potential. Thus the form factors are
much harder than those used by SA
(which show a decrease of up to 20 \%). Accordingly, the J\"ulich
$\pi N$ model does not show the deficiencies which, so far, have been
thought to be the main reason for the small contribution of the 
$\pi\pi$-exchange TBF in SA's work.

None the less it turned out that also the $\pi\pi$-exchange TBF
based on the $\pi N$ amplitude of the J\"ulich model is very small.
The contributions to the triton binding energy are in the order of a few
$keV$, which means comparable to the results obtained by Saito and Afnan.

A detailed analysis of the main differences between the
TM TBF and that derived from the J\"ulich $\pi N$ potential model
suggests that the differences in the contribution of the TBF to the
3N binding energy is due to the off-shell behaviour of the non-pole
$\pi N$ amplitude. In the J\"ulich
model the off-shell properties of the $\pi N$ amplitude are determined by
the dynamical ingredients of the model and the fit to the $\pi N$ data.
As a result, the off-shell properties of the amplitude are 
different in the individual $\pi N$ partial waves. In particular, the
$\pi N$ amplitudes that provide attractive contributions to the 
three-body binding ($P_{11}$, $P_{33}$) fall off relatively fast with 
increasing off-shell momentum while the repulsive $S_{31}$ partial wave
is enhanced. As a consequence, the total contribution of the TBF
is very small as a result of the cancellation effects. On the other hand,
in the $\pi N$ amplitude underlying the TM TBF, the off-shell
extrapolation is done in terms of a monopole form factor with a
cut-off mass of 800 MeV for all partial waves. 
Since this monopole form factor falls off much
slower then the $P_{11}$ and $P_{33}$ amplitudes of the J\"ulich $\pi N$
model, the corresponding attraction provided by the TM force is 
considerably enhanced. At the same time, the repulsion in the $S_{31}$
partial wave (and therefore any cancellation effects) is strongly
suppressed. These combined effects do indeed explain the 
two-orders-of-magnitude discrepancy in the resulting contribution to the 
triton binding energy as we have demonstrated in a numerical model study.

Naturally the question arises how realistic and well-defined the off-shell
properties of the J\"ulich $\pi N$ model are, especially in view of the
so-called quasi-potential ambiguity \cite{Friar}. This is a topic which
needs to be further investigated in the future, but it is certainly 
beyond the scope of the present paper. Here we only want to point to the
fact that rather different ansatzes for the $\pi N$ interaction (meson-exchange
and simple separable forms, respectively) 
lead to qualitatively similar off-shell features
and, in consequence, to similar results for the TBF. 
Qualitatively similar off-shell properties seem to be also predicted by
other $\pi N$ models - at least as far as we can judge from corresponding
publications \cite{PeJe,TaOh}.

With regard to the off-shell extrapolation used in the $\pi N$ amplitude
on which the TM TBF is based the situation is, in our opinion, much less
clear. First of all, 
we do not see any stringent physical reason for adopting the $\pi NN$
form factor for this purpose, especially because no $\pi NN$ vertex is
present at all in the non-nucleon-pole part of the $\pi N$ amplitude that
enters into the derivation of the TBF. Furthermore, the choice of having
the same off-shell properties in all $\pi N$ partial waves is also hard
to justify. Therefore the large contribution of the TM (and also the Brazil)
TBF to the triton binding energy of around 1 MeV - though certainly
desired by phenomenology - must be interpreted with caution. 
 
\bigskip

{\Large\bf Acknowledgements}

\smallskip

We would like to thank Prof. I.R. Afnan for valuable discussions
and for a careful reading of the manuscript. 
We acknowledge the hospitality of the RCNP in Osaka, Japan, where most
of the numerical calculations were carried out. 
This work was financially supported by the Deutsche Forschungsgemeinschaft
(Grant no. 447 AUS-113/3/0) and by the Japanese Society for the 
Promotion of Science.
 
\vfill \eject

\centerline{\bf Appendix: Separable expansion of a potential with two terms}
\bigskip

We consider to represent a potential $V$ which consists of
two terms, $V = V_1 + V_2$, in separable form.
At first we expand $V_2$ by means of the standard EST-method, cf. Eqs. 
(\ref{est1}) to (\ref{est4}). Then the t-matrix $\tilde T_2(E)$ obtained 
from the separable
representation $\tilde V_2$ agrees (on- as well as half-off-shell) with
$T_2(E)$ corresponding to $V_2$ at the choosen expansion energies
$E = E_i$, $i=1,...,N$. 

Next we expand the potential $V$ by assuming the following form
\begin{equation}
\tilde{V}(E) = |\tilde v_0\rangle \lambda_1(E) \langle\tilde v_0|
+ \tilde{V}_2 .
\label{eq5}\end{equation}
We determine the "form factor" $|\tilde v_0\rangle$ in such a way that
$\tilde V$ satisfies 
\begin{equation}
V |\psi_\varepsilon \rangle = \tilde{V}(\varepsilon) |\psi_\varepsilon\rangle,
\label{eq7}\end{equation}
where $|\psi_\varepsilon\rangle$ is a solution of the scattering equation, 
\begin{equation}
|\psi_\varepsilon\rangle = |k_\varepsilon\rangle
+ G_0 (\varepsilon) V |\psi_\varepsilon\rangle,
\label{eq8}\end{equation}
at a fixed predetermined energy $\varepsilon$. 
We want to emphasize that we may choose the energy $\varepsilon$ at which 
the potential $V$ is expanded to be different from any of the energies
$E_i$ chosen for the separable expansion of $V_2$ .
We see easily that Eq. (\ref{eq7}) is satisfied if we choose
$|\tilde v_0\rangle$ to be
\begin{equation}
|\tilde v_0\rangle = (V-\tilde{V}_2) \, |
\psi_\varepsilon \rangle ,
\label{eq9}\end{equation}
and $\lambda_1(\varepsilon)$ to be
\begin{eqnarray}
\lambda_1(\varepsilon) &=& {1\over
\langle\psi_\varepsilon | V - \tilde V_2 | \psi_\varepsilon \rangle} \\
&=&{1\over \langle\tilde v_0 | \psi_\varepsilon \rangle} \label{eq11}
\end{eqnarray}

Note that the (half-off-shell) t-matrix $\tilde{T}(E)$ obtained from 
$\tilde V(\varepsilon)$ is identical to $T(E)$ obtained from $V$  
at $E = \varepsilon$ because of condition (\ref{eq7}).

If $T(E)$ has a pole at $E_p$ we may choose the expansion energy 
$\varepsilon$ for the separable representation to be $\varepsilon = E_p$. 
The wave function at the pole, $| \psi_p \rangle$,
is a solution of the equation
\begin{equation}
|\psi_p\rangle = G_0(E_p) V \, |\psi_p \rangle \ .
\label{eq12}\end{equation}
The t-matrix $\tilde{T}(E)$ for the separable potential
$\tilde{V}(E)$ is given by 
\begin{equation}
\tilde T(E) = |\tilde v(E)\rangle \,
{1\over 1/\lambda_1(E) - \tilde\Sigma(E)} \, \langle \tilde v(E) |
+ \tilde T_2(E) \ , 
\label{eq13}\end{equation}
where
\begin{equation}
\tilde \Sigma(E) = \langle\tilde v_0 \, | \, G_0(E) \, | \,
\tilde v(E) \rangle,
\label{eq14}\end{equation}
and 
\begin{equation}
|\tilde v(E)\rangle = (1+ \tilde T_2(E) G_0(E)) \, |\tilde v_0 \rangle \ . 
\label{eq15}\end{equation}  
The "form factor" $|\tilde v_0 \rangle$ is defined by 
\begin{equation}
|\tilde v_0 \rangle = (V - \tilde V_2 )
\, |\psi_p \rangle .
\label{eq16}\end{equation}
Substituting Eqs. (\ref{eq15}) and (\ref{eq16}) into Eq. (\ref{eq14})
leads to 
\begin{equation}
\tilde \Sigma(E) =
\langle\tilde v_0 | (1 + G_0(E) \tilde T_2(E)) G_0(E) V
- G_0(E) \tilde T_2(E) | \psi_p \rangle \ . 
\label{eq17}\end{equation}
We can evaluate $\tilde\Sigma(E)$ at $E = E_p$
by using Eq. (\ref{eq12}), 
\begin{eqnarray}
\tilde\Sigma(E_p) &=& \langle\tilde f_0 | \psi_p \rangle
\nonumber \\
&=& {1\over \lambda_1(E_p)} \ , 
\label{eq18}\end{eqnarray}
where we have utilized, in addition, Eq. (\ref{eq11}).
By substituting this result into Eq. (\ref{eq13}), we see that
the t-matrix $\tilde T(E)$ obtained for
the separable potential $\tilde V$ has the same pole position
as the t-matrix of the original potentil $V$. 

Let us now assume that the first term of the original potential $V$ 
has a separable form, 
\begin{equation}
V_1(E) = |v_0\rangle \Lambda(E) \langle v_0 |,
\label{eq21}\end{equation}
where $\Lambda(E)$ is
\begin{equation}
\Lambda(E) = {1\over E - m_0}.
\label{eq22}\end{equation}
This is exactly the case for the (direct) nucleon pole contribution, 
Fig. \ref{fig:diags}(a). Then $|v_0 \rangle$ corresponds to the bare
$\pi NN$ vertex function and $m_0$ is the bare nucleon mass. 
The solution of the Lippmann-Schwinger equation for $V = V_1+V_2$ 
can then be written as 
\begin{equation}
T(E) = |v(E)\rangle {1\over 1/\Lambda(E) - \Sigma(E)}
\langle v(E) | + T_2(E),
\label{eq24}\end{equation}
where $T_2(E)$ is a solution of the scattering equation for the potential
$V_2$. The self-energy $\Sigma(E)$ is given by
\begin{equation}
\Sigma(E) = \langle v_0 | G_0(E) | v(E) \rangle,
\label{eq25}\end{equation}
and the dressed $\pi NN$ vertex function, $|v(E) \rangle$, by 
\begin{equation}
|v(E) \rangle = (1 + T_2(E) G_0(E)) \, |v_0 \rangle . 
\label{eq26}\end{equation}
Assuming that the t-matrix of Eq. (\ref{eq24}) has a pole at $E= m_N$,
$\Sigma(m_N)$ is evaluated as
\begin{eqnarray}
\Sigma(m_N) &=& {1\over\Lambda(m_N)} \nonumber \\
&=& m_N - m_0 \ . \label{eq31}
\end{eqnarray}
Also, the pole part of the t-matrix can be written in the form,
\begin{equation}
T(E) = |v(E)\rangle {1\over 1/\Lambda(E) - \Sigma(E)}
\langle v(E) | \ = \  
|v^R(E)\rangle {1\over {E - m_N}} \langle v^R(E) | \ ,    
\label{eq24a}\end{equation}
which defines the renormalized $\pi NN$ vertex function $|v^R(E)\rangle$.
At the pole $|v^R(E)\rangle$ is given by (cf., e.g., Ref. \cite{SA95})
\begin{equation}
|v^R(m_N)\rangle = {|v(m_N)\rangle \over (1 - \Sigma'(m_N))^{1/2}},
\label{eq28}\end{equation}
where
\begin{equation}
\Sigma'(m_N) = - \langle v(m_N) | G^2_0(m_N) | v(m_N) \rangle \ .
\label{eq29}\end{equation}
In the last step we have assumed that $V_2$ does not dependent on the
energy $E$. If it does (like in our case) then $\Sigma'$ is determined
by 
\begin{equation}
\Sigma'(m_N) = {\partial \over \partial E}
 \langle v_0 | G_0(E) | v(E) \rangle \ |_{E=m_N} \ .
\label{eq29a}\end{equation}
Furthermore, we note that the renormalized vertex function is related
to the $\pi NN$ coupling constant $f_{\pi NN}$ and 
the $\pi NN$ form factor $F_{\pi NN}$ by
\begin{equation}
f_{\pi NN}  F_{\pi NN}({\bf q}^2) a(q) \ = \
\langle q  | v^R(m_N) \rangle \ .
\label{eq29b}\end{equation}
where $a(q)$ is a kinematical factor depending on the particular 
($ps$ or $pv$) $\pi NN$ coupling (cf. Eqs. (3) and
(16) of Ref. \cite{CSch2}). 

The wave function at the pole energy is obtained by solving Eq. (\ref{eq12})
\begin{equation}
|\psi_p \rangle = G_0(m_N) \, |v(m_N) \rangle.
\label{eq30}\end{equation}
By substituing Eq. (\ref{eq30}) into the $lhs$ of Eq. (\ref{eq12}),
we get another relation between $|v(m_N)\rangle$ and
$|\psi_p\rangle$, 
\begin{equation}
|v(m_N)\rangle = V(m_N) | \psi_p \rangle .
\label{eq32}\end{equation}

Now we expand the potential model $V(E)$ using the method described above. 
Then the resulting t-matrix obtained from the separable
representation has the same pole position as the original interaction
model. We will examine whether also the $\pi NN$ coupling constant and
form factor determined from the separable representation agree with the 
ones of the original interaction.

At first, we derive a relation between $|\tilde v(m_N) \rangle$
and $|v(m_N)\rangle$. Using Eqs. (\ref{eq15}), (\ref{eq16}), and
finally (\ref{eq32}) we obtain 
\begin{eqnarray}
|\tilde v(m_N) \rangle &=& (1 + \tilde T_2(m_N) G_0(m_N) )
(V(m_N) - \tilde V_2) | \psi_p \rangle \nonumber \\
&=& V(m_N) |\psi_p \rangle \nonumber \\
&=& |v(m_N) \rangle . \label{eq33}\end{eqnarray}
Eq. (\ref{eq33}) implies that the momentum dependence of the $\pi NN$ vertex
function obtained from the separable representation is identical to the one 
of the original potential.

The renormalized $\pi NN$ vertex function at the pole, 
$|\tilde v^R(m_N)\rangle$, for 
the separable representation $\tilde V$ can be calculated in the same 
way as for the original potential. It is obtained by
placing tildes over all quantities in Eq. (\ref{eq28})
and (\ref{eq29}): 
\begin{eqnarray}
|\tilde v^R(m_N)\rangle &=& {|\tilde v(m_N)\rangle \over 
(1 - \tilde \Sigma'(m_N))^{1/2}},
\label{eq34} \\
\tilde\Sigma'(m_N) &=& - \langle \tilde v(m_N)|G^2_0(m_N)|
\tilde v(m_N) \rangle. \label{eq35}
\end{eqnarray}
Because of Eq. (\ref{eq33}) we get
\begin{equation}
\tilde\Sigma'(m_N) = \Sigma'(m_N) \ .
\label{eq36}\end{equation}
Finally, by substituting Eqs. (\ref{eq33}) and (\ref{eq36}) into Eq. 
(\ref{eq34}) it follows that 
\begin{equation}
| \tilde v^R({m_N}) \rangle  = | v^R({m_N}) \rangle \ , 
\label{eq37}\end{equation}
and consequently that $\tilde f_{\pi NN} = f_{\pi NN}$ and 
$\tilde F_{\pi NN}({\bf q}^2) = F_{\pi NN}({\bf q}^2)$. 

In the above discussion we have not specified the form of
$\lambda_1(E)$, since we needed only the value of $\lambda_1(E)$
at the energy $E=m_N$ to get the correct half-off shell t-matrix.
If we assume $\lambda_1(E)$ to be of the form 
$1/\lambda_1(E) = E - \tilde m_0$, then $\tilde m_0$ can be
determined by Eq. (\ref{eq11}) (or Eq. (\ref{eq18})) to be 
\begin{equation}
\tilde m_0 = m_N - \langle\psi_p | V - \tilde V_2 | \psi_p \rangle.
\label{eq38}\end{equation}

Therefore, by following the outlined procedure
it is possible to construct a separable representation where
(i) the half-off-shell behavior of the non-pole part of the t-matrix 
(at selected energies), 
(ii) the  pole position, (iii) the $\pi NN$ form factor, and
(iv) the value of the $\pi NN$ coupling
constant are the same as in the original interaction model.

\vfill \eject

\vfill \eject

\begin{table}[htbp]
\begin{center}
\caption{The contribution of the $\pi\pi$-exchange three-body force to the
triton binding energy in $keV$ for the $\pi N$ models 1, 2 and 2'
of the J\"ulich group \protect\cite{CSch1,CSch2} and 
$PJ$ of Ref. \protect\cite{SA95}. The $\pi NN$ form factor $F_{\pi NN}
(q^2=0)$ determined consistenly from these models is given in the second
line. }
\label{rstt11}
\begin{tabular}{crrrr}
\hline\hline
  & model 1' & model 2 & model 2'& $PJ$  \\
\hline
$F_{\pi NN}(0)$ &
           0.934 & 0.925 & 0.962 & 0.811 \\
\hline\hline
$S_{11}$ &  -3.8 &  -2.8 &  -2.3 &  -4.8 \\
$S_{31}$ &  35.1 &  64.4 &  51.7 &  26.4 \\
$P_{11}$ & -11.8 & -26.4 & -21.3 &  -8.8 \\
$P_{31}$ &   1.8 &   9.0 &   6.4 &  -3.6 \\
$P_{13}$ &  -2.2 &  -0.9 &  -1.1 &   4.5 \\
$P_{33}$ & -17.1 & -31.6 & -24.2 & -16.0 \\
\hline
$Total$  &   2.0 &  11.7 &   9.2 &  -2.3 \\
\hline\hline
\end{tabular}\end{center}
\end{table}

\begin{table}[htbp]
\begin{center}
\caption{The effect of choosing different expansion energies 
for the rank-1 separable representation in the EST expansion.
Columns labelled with $E_i < E_{th}$ correspond to our standard choice
of $E_i = 930$ MeV. The results given in the 
columns labelled with $E_i > E_{th}$ 
are obtained for the value of $1077$ MeV for $S_{11}$,
$1000$ MeV for $S_{31}$ and $1100$ MeV for the other partial waves.
All binding energies are given in $keV$.}
\label{rstt2}
\begin{tabular}{crrrr}
\hline
&\multicolumn{2}{c}{model 1}&\multicolumn{2}{c}{model 2'}\\
& $E_i > E_{th}$ & $E_i < E_{th}$ & $E_i > E_{th}$ &
$E_i < E_{th}$ \\
\hline
$S_{11}$&  -5.4 &  -3.8 &  -5.5 &  -2.3 \\
$S_{31}$&  37.5 &  35.1 &  59.0 &  51.7 \\
$P_{11}$& -11.8 & -11.8 & -20.6 & -21.3 \\
$P_{31}$&  -1.0 &   1.8 &   0.3 &   6.4 \\
$P_{13}$&  -2.1 &  -2.2 &  -1.5 &  -1.1 \\
$P_{33}$& -14.7 & -17.1 & -20.5 & -24.2 \\
\hline
 Total &   2.5 &   2.0 &  11.2 &   9.2 \\
\hline
\end{tabular}
\end{center}
\end{table}

\begin{table}[htbp]
\begin{center}
\caption{The effect of removing the energy dependence in the $\pi N$
potential (i), and in addition in the $\pi NN$ vertex function and the
$\pi N$ ampliude (ii). All results are obtained by using model 2'. 
The binding energies are in $keV$. }  
\label{dsct3}
\begin{tabular}{crrr}
\hline
    & exact   &  (i)  &  (ii) \\
\hline
 $S_{11}$ &  -2.3 &  -3.4 &  -3.9 \\
 $S_{31}$ &  51.7 &  76.5 &  74.7 \\
 $P_{11}$ & -21.3 & -33.1 & -39.9 \\
 $P_{31}$ &   6.4 &  13.7 &  14.4 \\
 $P_{13}$ &  -1.1 &   0.9 &   1.1 \\
 $P_{33}$ & -24.2 & -41.9 & -45.7 \\
\hline
 $Total$  &   9.2 &  12.7 &  -0.7 \\
\hline
\end{tabular}\end{center}
\end{table}

\begin{table}[htbp]
\begin{center}
\caption{The effect of replacing the off-shell properties of our $\pi N$
amplitude by monopole type functions (cf. Eq. \ref{disc2}) with different 
cutoff masses $\Lambda$. 
Starting point is model 2'. The binding energies are in $keV$. }  
\label{dsct4}
\begin{tabular}{crrrr}
\hline
 $\Lambda (MeV) $ & 200 & 400 & 600 & 800 \\
\hline
 $S_{11}$ &  -0.7 &  -4.8 &   -8.6 &  -11.0 \\
 $S_{31}$ &   0.4 &   7.0 &   13.4 &   17.5 \\
 $P_{11}$ &  -7.1 & -28.5 &  -63.6 & -100.2 \\
 $P_{31}$ &  -2.1 &   3.2 &   12.4 &   19.6 \\
 $P_{13}$ &  -1.1 &   1.1 &    5.6 &    9.3 \\
 $P_{33}$ &  -0.5 & -48.7 & -129.7 & -202.7 \\
\hline
 $Total$  & -11.1 & -70.7 & -170.5 & -267.5 \\
\hline
\end{tabular}\end{center}
\end{table}

\vfill \eject

\begin{figure}
\caption{The contribution to the three-nucleon force.}
\label{ftbf}
\end{figure}

\begin{figure}
\caption{Diagrams included in the $\pi N$ potential.}
\label{fig:diags}
\end{figure}

\begin{figure}
\caption{Comparison of the background isoscalar $\pi N$
amplitude $\bar F^+(0,t)$. The solid (long-dashed) 
lines are obtained from the J\"ulich $\pi N$ model 1 (2) of Ref. 
\protect\cite{CSch1}. The dash-dotted (short-dashed) lines are 
the predictions of the amplitudes that form the basis of the 
Tucson-Melbourne (Brazil) $\pi \pi$-exchange three-body force. 
Empirical values at the Cheng-Dashen (CD) point, $t=m_\pi^2$ and
$t=0$ are taken from Ref. \protect\cite{Hoe}.}  
\label{fig:cdp}
\end{figure}

\begin{figure}
\caption{$\pi NN$ vertex functions
as function of the pion momentum in the $\pi N$ c.m. system.
The solid (long-dashed) line denotes the prediction resulting from
model 1 (2) of Ref. \protect\cite{CSch1},
the dash-dotted (short-dashed) line denotes the prediction of
model 1' (2') of Ref. \protect\cite{CSch2}, the dotted line shows
the result of model $PJ$ of Ref. \protect\cite{McLeod}.}
\label{fig:fofa}
\end{figure}

\begin{figure}
\caption{Off-shell $\pi N$ transition amplitude $T_\alpha (q,q';Z)$ for
$q$ = $q'$  = 130 MeV/c as a function of the total energy $Z$.
The full line is the result of model 2' of Ref. \protect\cite{CSch2}
whereas the dashed line corresponds to the rank-1 separable
representation described in the text.}
\label{fig:pin1}
\end{figure}

\begin{figure}
\caption{Off-shell $\pi N$ transition amplitude $T_\alpha (q,q';Z)$ for
$q'$ = 130 MeV/c at three different energies $Z$ as a function
of the off-shell momentum $q$. Same description as in 
Fig. \protect\ref{fig:pin1}.} 
\label{fig:pin2}
\end{figure}

\begin{figure}
\caption{Illustration of the three-body force. See the text
for the definitions of functions, momenta and energies.}
\label{figtbf1}
\end{figure}

\begin{figure}
\caption{Comparison of the monopole type function $F(q) = \Lambda^2
/ (\Lambda^2 + q^2)$ with $\Lambda = 800$ MeV with the 
off-energy shell behaviour of our model 2' for various $\pi N$ partial waves.
$f_\alpha(q)$ is defined in Eq. (\protect\ref{disc1}).}
\label{figtbf2}
\end{figure}


\begin{thebibliography}{1}
\bibitem{Com}
Note that "form factors" in the context of separable interactions have 
to be distinguished from the vertex form factors of meson-exchange models. 
These are different objects! However, in case of pole diagrams 
there is a one-to-one correspondence between the "form factor" of a 
separable potential and the
vertex function, as expressed in Eq. (\ref{eq21}) of the Appendix. In order
to minimize confusion we will always use quotation marks if we speak about
"form factors" of separable potentials.
\end{thebibliography}
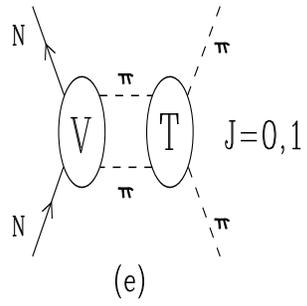
\end{document}